\documentclass[twocolumn,twoside,dvipdfm]{IEEEtran}
\usepackage{}            
\usepackage{amsmath}%
\usepackage{MnSymbol}%
\usepackage{wasysym}%
\usepackage{color, soul}
\usepackage[pdftex]{graphicx}

\newtheorem{definition}{Definition}
\newtheorem{proposition}{Proposition}
\newtheorem{lemma}{Lemma}

\newtheorem{example}{Example}

\ifCLASSINFOpdf
\else
\fi
\usepackage{algorithmic}
\usepackage{algorithm}
\begin{document}
%
\title{On the Successive Cancellation Decoding of Polar Codes with Arbitrary Linear Binary Kernels}
%
%
%

\author{Zhiliang~Huang,
        Shiyi~Zhang,~Feiyan~Zhang,~Chunjiang~Duanmu,~and~Ming~Chen 
\thanks{Zhiliang Huang, Shiyi Zhang, Feiyan Zhang, and Chunjiang Duanmu are with the School of Mathematics, Physics and Information Engineering, Zhejiang Normal University, Jinhua, 321004, China (e-mail: zlhuang, syzhang, zhangfy, duanmu@zjnu.edu.cn).}
\thanks{Ming Chen is with the School of Information Science and Engineering, the National Mobile Communications Research Laboratory, Southeast University, Nanjing, 210096, China (e-mail: chenming@seu.edu.cn).}
}

\maketitle

\begin{abstract}
A method for efficiently successive cancellation (SC) decoding of polar codes with high-dimensional linear binary kernels (HDLBK) is presented and analyzed. We devise a $l$-expressions method which can obtain simplified recursive formulas of SC decoder in likelihood ratio form for arbitrary linear binary kernels to reduce the complexity of corresponding SC decoder.
By considering the bit-channel transition probabilities $W_{G}^{(\cdot)}(\cdot|0)$ and $W_{G}^{(\cdot)}(\cdot|1)$ separately, a $W$-expressions method is proposed to further reduce the complexity of HDLBK based SC decoder. For a $m\times m$ binary kernel, the complexity of straightforward SC decoder is $O(2^{m}N\log N)$. With $W$-expressions, we reduce the complexity of straightforward SC decoder to $O(m^{2}N\log N)$ when $m\leq 16$. Simulation results show that $16\times16$ kernel polar codes offer significant advantages in terms of error performances compared with $2\times2$ kernel polar codes under SC and list SC decoders.
\end{abstract}


%
\IEEEpeerreviewmaketitle

\section{Introduction}
\IEEEPARstart{P}{olar} codes were introduced by Ar{\i}kan \cite{Arikan} as the first family of capacity achieving codes with explicit construction method and low encoding/decoding complexities for the class of binary input discrete memoryless channels (B-DMCs). Ar{\i}kan's original polar codes is based on the kernel matrix
$G_{2}=\left(
        \begin{array}{cc}
          1 & 0 \\
          1 & 1 \\
        \end{array}
      \right)$
and its $n$th Kronecker power $G_{2}^{\otimes n}$ corresponding to a linear code with block length $N=2^{n}$. With Ar{\i}kan's $2\times 2$ kernel, it was shown in \cite{rate} the probability of block error under successive cancellation (SC) decoding is $o(2^{-2^{n\beta}})$ with $\beta=0.5$. It was conjectured in \cite{Arikan} that channel polarization is a general phenomenon and it was shown in \cite{exponent} that the probability of block error under SC decoding is $o(2^{-m^{n\beta}})$ for a general kernel $G_{m}$ with size $m\times m$. $\beta$ is called \emph{exponent} of the kernel and can exceed $0.5$ for large $m$ \cite{exponent}.

\begin{figure}[htb]
\centering
\includegraphics[width=7.0cm,height=7.0cm]{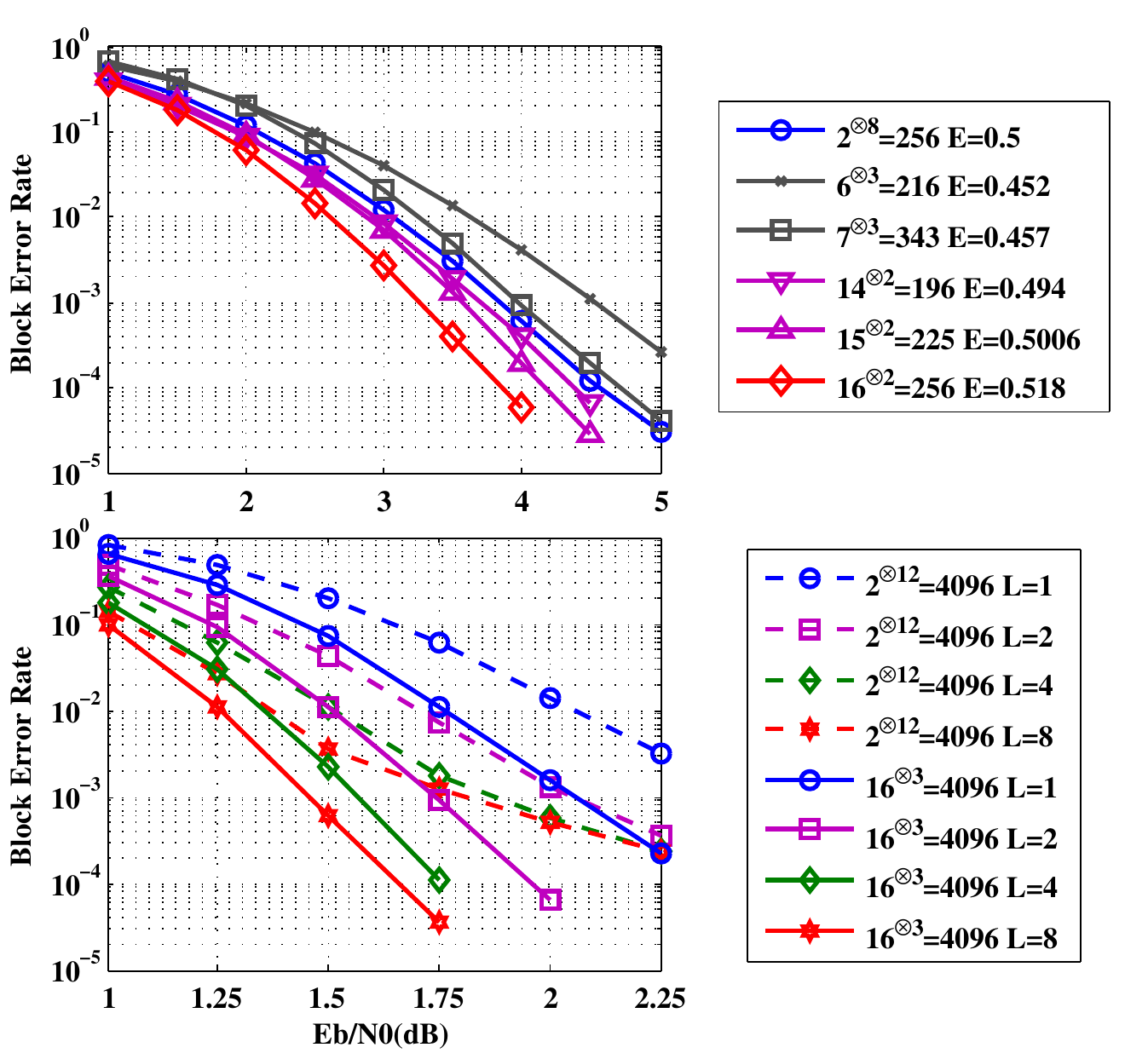}
\caption{ The code rate is 0.5. Top: SC decoding performance of polar codes with kernels $G_{2}$, $G_{6}$, $G_{7}$, $G_{14}$, $G_{15}$, $G_{16}$ on the BPSK-modulated Gaussian channel. All codes are constructed using Gaussian Approximation method \cite{GA} at Eb/N0$=2$dB. Bottom: List SC \cite{List} decoding performance of polar codes with kernels $G_{2}$ and $G_{16}$. $G_{2}^{\otimes 12}$ code is constructed via the method proposed in \cite{DE&UP} at Eb/N0$=2$dB  and $G_{16}^{\otimes 3}$ code is based on Monte Carlo method proposed in [1, Sec. IX] at Eb/N0$=2$dB.}
\label{fig2-16}
\end{figure}

Based on the above, many researchers had constructed high-dimensional kernels with large exponents. Based on BCH codes, Korada \emph{et al.} \cite{exponent} provided a construction of binary kernels with large exponent. Mori and Tanaka \cite{non-binary} proposed a construction of non-binary kernels with large exponent based on Reed-Solomon codes. In \cite{decomposition}, code decompositions were used to design good linear and nonlinear binary kernels. In \cite{Hisien-ping}, constructions were presented for kernels with maximum exponents up to dimension $16$.

However, it was pointed out in \cite{exponent} that the complexity of straightforward SC decoder for $G_{m}$ polar codes behaved like $O(2^{m}N\log_{m}^{N})$. So it's not practical for high-dimensional kernels such as $m=16$. At present, to the best of our knowledge, there is no efficient SC decoding of large dimension kernels, although exponent's definition is base on SC decoding. In \cite{Bonik} and \cite{Wang}, they tried to generalize the idea of $G_{2}$ SC decoding to high-dimensional binary kernels. But their methods can only work on kernels with very small dimension because their methods need a tree structure for \emph{bit-channel graph} \cite{Bonik} and it's not true for large dimension kernels even with $m=6$.

In this paper, we propose a low complexity SC decoder for arbitrary binary linear kernels. For $G_{2}$, it has $l_{2}^{(1)}=l_{1}\diamond l_{2}
\footnote{$l_{1}\diamond l_{2}=(l_{1}l_{2}+1)/(l_{1}+l_{2})$},l_{2}^{(2)}=l_{1}^{1-2u_{1}}l_{2}$ which are called the $l$-expressions in this paper leading to a low complexity SC decoding of $G_{2}$ polar codes \cite{Arikan}. Our basic idea, like $G_{2}$, is to obtain $l$-expressions for arbitrary binary linear kernel $G_{m}$.

Fig. \ref{fig2-16} shows the error performance of polar codes with different kernels under SC and list SC (LSC) decoding through binary-input additive white gaussian noise (AWGN) channel. It can be seen that error performance of polar codes are almost decided by the exponent of kernels. $G_{6}$ and $G_{7}$ have smaller exponent than $G_{2}$ and the error performance of polar codes with $G_{6}$ and $G_{7}$ is worse than $G_{2}$ even with longer block length. For kernels with close exponents, high-dimensional kernel polar codes have better error performance than $G_{2}$ kernel polar codes even with shorter block length such as $G_{14}^{\otimes2}$ and $G_{15}^{\otimes2}$. $G_{16}$ polar codes achieves significant error performance gains than $G_{2}$ polar codes under SC and LSC decoding, although $G_{16}$'s exponent is a little bigger than $G_{2}$.

Next, we use an example to show our main idea.
\begin{example}[$l$-expressions for an optimal $G_{6}$ kernel.]
\label{G6}
\[
G_{6}=
\left(
        \begin{array}{cccccc}
          1 & 0 & 0 & 0 & 0 & 0 \\
          1 & 1 & 0 & 0 & 0 & 0 \\
          1 & 0 & 1 & 0 & 0 & 0 \\
          1 & 0 & 0 & 1 & 0 & 0 \\
          1 & 1 & 1 & 0 & 1 & 0 \\
          1 & 1 & 0 & 1 & 0 & 1 \\
        \end{array}
      \right)
\]
The $l$-expressions for this kernel:
\begin{align*}
l_{6}^{(1)} & =l_{1}\diamond l_{2}\diamond l_{3}\diamond l_{4}\diamond l_{5}\diamond l_{6}\\
l_{6}^{(2)} & =(l_{1}\diamond l_{3}\diamond l_{4})(l_{2}\diamond l_{5}\diamond l_{6})\\
l_{6}^{(3)} & =(l_{1}\diamond l_{4}l_{3})\diamond (l_{2}\diamond l_{6}l_{5})\\
l_{6}^{(4)} & =l_{1}\diamond(l_{2}(l_{3}l_{5})\diamond(l_{4}l_{6}))\boxtimes l_{4}\diamond((l_{1}^{-1}l_{2})\diamond(l_{3}l_{5})l_{6})\\
l_{6}^{(5)} & =(l_{1} l_{2})\diamond(l_{4} l_{6})(l_{3}l_{5})\\
l_{6}^{(6)} & =l_{1}l_{2} l_{4} l_{6}
\end{align*}
\end{example}
In Example \ref{G6}, we give $l$-expressions for a $6\times 6$ optimal kernel. For the above $G_{6}$ kernel, the complexity of straightforward SC decoder is $O(2^{6}N\log N)$. With these $l$-expressions, we reduce the complexity to $O(7N\log N)$ where $7$ is the \emph{length} (defined later on) of $l$-expressions.

The above $G_{6}$ kernel is given in \cite{Hisien-ping} and \emph{optimal} means it has maximum exponent among all $6\times 6$ kernels. $l_{1},\cdots, l_{6}$ defined by $l_{i}=W(y_{i}|0)/W(y_{i}|1)$ ($W$ is the channel) are channel likelihood ratios and $l_{6}^{(1)},\cdots, l_{6}^{(6)}$ are bit-channels's likelihood ratios. $l_{i}l_{j}$ means $l_{i}\times l_{j}$. For $l_{6}^{(4)}$, two parts which are connected by $\boxtimes$ are called \emph{sub-expressions}. $\boxtimes$ is the same like $\times$ and it is specially used for separating sub-expressions. Three operators's priority is $\boxtimes<\times<\diamond$. Then $l_{i}\diamond l_{j}l_{k}=(l_{i}\diamond l_{j})\times l_{k}$. In Example \ref{G6}, we omit the influence of known values $u_{1}^{i-1}$ for $l_{6}^{(i)}$. For example, $l_{6}^{(3)}=(l_{1}^{(1-2(u_{1}+u_{2}))}\diamond l_{4}l_{3})\diamond (l_{2}^{(1-2u_{2})}\diamond l_{6}l_{5})$ and it will be explained in section III. C.



By our analysis, $l$-expressions based SC decoder is good for medium kernel size such as $m\leq 10$. However, it become impractical for lager kernel size such as $m=16$. So, similar to $l$-expressions, we propose a $W$-expressions method to further reduce the complexity of SC decoder for larger dimension kernels by considering bit-channel transition probabilities $W_{G}^{(\cdot)}(\cdot|0)$ and $W_{G}^{(\cdot)}(\cdot|1)$ separately. Our main achievement is: \emph{Using $W$-expressions method, we show that the complexity of $G_{m}$ SC decoder is $O(m^{2}N\log N)$ for optimal kernels given in \cite{Hisien-ping} when $m\leq 16$.}

The rest of the paper is organized as follows. In section II, we introduce the basic definitions and point out our basic task. In section III, we give details how to get $l$-expressions for an arbitrary binary kernel matrix. In section IV, similar to $l$-expressions, we present a $W$-expressions method to further reduce the complexity of SC decoder with high dimensional kernel. Also, we give complexity analyses of $l$-expressions and $W$-expressions based SC decoder in this section. Construction methods of polar codes with high dimensional kernel are presented in section V.

\section{Preliminaries}
\subsection{Notations}
We write $W:\{0,1\}\rightarrow \mathcal{Y}$ to denote a B-DMC channel with input alphabet $\{0,1\}$, output alphabet $\mathcal{Y}$, and transition probabilities $W(y|x), x\in \{0,1\}, y\in \mathcal{Y}$. We use the notation $a_{1}^{N}$ for denoting a row vector $(a_{1},\cdots,a_{N})$.
For a general kernel matrix $G_{m}$ (all kernels used in this paper are linear kernels given in \cite{Hisien-ping}), $W_{G_{m}}:\{0,1\}^{m}\rightarrow \mathcal{Y}^{m}$ is defined by
\begin{equation}\label{eq1}
W_{G_{m}}(y_{1}^{m}|u_{1}^{m})\triangleq \prod_{i=1}^{l}W(y_{i}|(u_{1}^{m}G_{m})_{i}).
\end{equation}
Then, bit-channels $W_{G_{m}}^{(i)}:\{0,1\}\rightarrow \mathcal{Y}^{m}\times \{0,1\}^{i-1},1\leq i\leq m$ is defined by
\begin{equation}\label{def_bitchannel}
W_{G_{m}}^{(i)}(y_{1}^{m},u_{1}^{i-1}|u_{i})\triangleq \frac{1}{2^{m-1}}\sum_{u_{i+1}^{m}}W_{G_{m}}(y_{1}^{m}|u_{1}^{m}).
\end{equation}
For SC decoding, the basic task is to calculate following values
\begin{multline}\label{def}
l_{m}^{(i)}=\frac{W_{G_{m}}^{(i)}(y_{1}^{m},u_{1}^{i-1}|u_{i}=0)}{W_{G_{m}}^{(i)}(y_{1}^{m},u_{1}^{i-1}|u_{i}=1)}\\
  =\frac{\sum_{u_{i+1}^{m}}W(y_{1}|(u_{1,u_{i}=0}^{m}G_{m})_{1})\cdots W(y_{m}|(u_{1,u_{i}=0}^{m}G_{m})_{m})}
   {\sum_{u_{i+1}^{m}}W(y_{1}|(u_{1,u_{i}=1}^{m}G_{m})_{1})\cdots W(y_{m}|(u_{1,u_{i}=1}^{m}G_{m})_{m})}
\end{multline}
where $u_{1,u_{i}=0}^{m}$ means $(u_{1}^{i-1},u_{i}=0,u_{i+1}^{m})$.

In order to facilitate notation. We use following simple notation instead of (3)
\begin{equation}\label{SN_def_bit}
l_{m}^{(i)}=\frac{\sum_{u_{i+1}^{m}}W(y_{1}|\textbf{u}_{1})\cdots W(y_{i}|\textbf{u}_{i})\cdots W(y_{m}|\textbf{u}_{m})}
                {\sum_{u_{i+1}^{m}}W(y_{1}|\overline{\textbf{u}}_{1})\cdots W(y_{i}|\overline{\textbf{u}}_{i})\cdots W(y_{m}|\textbf{u}_{m})}
\end{equation}
where $\textbf{u}_{k}=(u_{1,u_{i}=0}^{m}G_{m})_{k}, k=1,\cdots,m$, $\overline{\textbf{u}}_{k}=\textbf{u}_{k}+1$. Let $g_{ik}$ denote the element of $G_{m}$ in the $i$th row and $k$th column. In the denominator of (\ref{SN_def_bit}), if $g_{ik}=1$, $(u_{1,u_{i}=1}^{m}G_{m})_{k}=\overline{\textbf{u}}_{k}$; otherwise $(u_{1,u_{i}=1}^{m}G_{m})_{k}=\textbf{u}_{k}$.

\newcounter{mytempeqncnt}
\begin{figure*}[tbp]
\begin{example}
\label{l-ex_main_idea}
\normalsize
\setcounter{mytempeqncnt}{\value{equation}}
\begin{align}
\label{eqn_dbl_x}
  l_{6}^{(4)} & = \frac{\sum_{u_{5}^{6}}W(y_{1}|u_{5}+u_{6})W(y_{2}|u_{5}+u_{6})W(y_{3}|u_{5})W(y_{4}|u_{6})W(y_{5}|u_{5})W(y_{6}|u_{6})}
     {\sum_{u_{5}^{6}}W(y_{1}|\overline{u_{5}+u_{6}})W(y_{2}|u_{5}+u_{6})W(y_{3}|u_{5})W(y_{4}|\overline{u_{6}})W(y_{5}|u_{5})W(y_{6}|u_{6})} \\
 \label{extend1}
   & =\frac{\sum_{u_{5}^{6}}W(y_{1}|u_{5}+u_{6})W(y_{2}|u_{5}+u_{6})W(y_{3},y_{5}|u_{5})W(y_{4}|u_{6})W(y_{6}|u_{6})}
     {\sum_{u_{5}^{6}}W(y_{1}|\overline{u_{5}+u_{6}})W(y_{2}|u_{5}+u_{6})W(y_{3},y_{5}|u_{5})W(y_{4}|u_{6})W(y_{6}|u_{6})} \\
 \label{extend2}
   &\boxtimes \frac{\sum_{u_{5}^{6}}W(y_{1}|\overline{u_{5}+u_{6}})W(y_{2}|u_{5}+u_{6})W(y_{3},y_{5}|u_{5})W(y_{4}|u_{6})W(y_{6}|u_{6})}
     {\sum_{u_{5}^{6}}W(y_{1}|\overline{u_{5}+u_{6}})W(y_{2}|u_{5}+u_{6})W(y_{3},y_{5}|u_{5})W(y_{4}|\overline{u_{6}})W(y_{6}|u_{6})}  \\
   \label{fundenmental1}
   &=l_{1}\diamond \frac{\sum_{u_{5}^{6}}W(y_{2}|u_{5}+u_{6})W(y_{3},y_{5}|u_{5})W(y_{4},y_{6}|u_{6})\textbf{1}_{u_{5}+u_{6}=0}}
     {\sum_{u_{5}^{6}}W(y_{2}|u_{5}+u_{6})W(y_{3},y_{5}|u_{5})W(y_{4},y_{6}|u_{6})\textbf{1}_{u_{5}+u_{6}=1}}
     \boxtimes
     \frac{\sum_{u_{6}}W(y_{1}^{2},y_{3},y_{5}|u_{6})W(y_{4}|u_{6})W(y_{6}|u_{6})}
     {\sum_{u_{6}}W(y_{1}^{2},y_{3},y_{5}|u_{6})W(y_{4}|\overline{u_{6}})W(y_{6}|u_{6})}  \\
     \label{fundenmental2}
   &=l_{1}\diamond \left(l_{2}\frac{\sum_{u_{6}}W(y_{3},y_{5}|u_{6})W(y_{4},y_{6}|u_{6})}
     {\sum_{u_{6}}W(y_{3},y_{5}|\overline{u_{6}})W(y_{4},y_{6}|u_{6})}\right)
     \boxtimes
     l_{4}\diamond \frac{\sum_{u_{6}}W(y_{1}^{2},y_{3},y_{5},y_{6}|u_{6})\textbf{1}_{u_{6}=0}}
     {\sum_{u_{6}}W(y_{1}^{2},y_{3},y_{5},y_{6}|u_{6})\textbf{1}_{u_{6}=1}} \\
      \label{l-expressions}
   &=l_{1}\diamond(l_{2}(l_{3}l_{5})\diamond(l_{4}l_{6}))\boxtimes l_{4}\diamond((l_{1}^{-1}l_{2})\diamond(l_{3}l_{5})l_{6}) \\
   \label{final-l-expressions}
   &=l_{1}^{(1-2(u_{1}+u_{2}+u_{3}))}\diamond(l_{2}^{(1-2u_{2})}(l_{3}^{(1-2u_{3})}l_{5})\diamond(l_{4}l_{6}))\boxtimes l_{4}\diamond((l_{1}^{-(1-2(u_{1}+u_{2}+u_{3}))}l_{2}^{(1-2u_{2})})\diamond(l_{3}^{(1-2u_{3})}l_{5})l_{6})
\end{align}
\hrulefill
\vspace*{4pt}
\end{example}
\end{figure*}

In (\ref{SN_def_bit}), we see \emph{channel expression} $W(y_{1}|\textbf{u}_{1})$ in numerator is different from $W(y_{1}|\overline{\textbf{u}}_{1})$ in denominator. We call this as one \emph{difference} for $l_{m}^{(i)}$.


It should be noticed that $\textbf{u}_{i}$ views as a expression of binary variables, not a value in our algorithm. For example, $\textbf{u}_{1}=u_{4}+u_{5}+u_{6}$ in the definition of $l_{6}^{(3)}$ by (\ref{def}). Let $\textbf{s}_{i}$ denote the set of variables contained in $\textbf{u}_{i}$. So $\textbf{s}_{1}=\{u_{4},u_{5},u_{6}\}$ in the previous example. $\textbf{u}_{i}\cap\textbf{u}_{j}=\emptyset$ means $\textbf{s}_{i}\cap\textbf{s}_{j}=\emptyset$. $\textbf{u}_{i}\subset\textbf{u}_{j}$ means $\textbf{s}_{i}\subset\textbf{s}_{j}$.

All of operations in this paper will be over GF(2). So, if $\textbf{u}_{1}=u_{4}+u_{5}$ and $\textbf{u}_{2}=u_{5}+u_{6}$, $\textbf{u}_{1}+\textbf{u}_{2}=u_{4}+u_{6}$.

We write $\textbf{1}_{\textbf{u}_{1}=0}$ to denote the indicator function of equation $\textbf{u}_{1}=0$; thus, $\textbf{1}_{\textbf{u}_{1}=0}$ equals $1$ if $\textbf{u}_{1}=0$ and $0$ otherwise.

In Example \ref{G6}, $l_{6}^{(4)}$ is connected by two sub-expressions with $\boxtimes$. We call the \emph{length} of $l_{6}^{(4)}$ is $2$. And other $l$-expressions's lengths are all $1$. So the total length is $7$ for this example.

\subsection{Basic task}
By using (\ref{def}), the total computational cost of $l_{m}^{(1)},\cdots,l_{m}^{(m)}$ is $O(m2^{m})$. We call these calculations as \emph{inside kernel calculation}. A polar code defined by $G_{m}^{\otimes n}$ with block length $N=m^{n}$ needs to recursively implement $N\log_{m}^{N}/m$ times of \emph{inside kernel calculation} for SC decoding. So the complexity of SC decoding behaves like $O(2^{m}N\log_{m}^{N})$ for a general kernel matrix $G_{m}$. It grows exponentially with the kernel size. So it is not practical for large kernel size such as $m=16$. Therefore, our basic task is to reduce the computational cost of (\ref{def}).

\section{Bit-channel Likelihood Expressions for $G_{m}$}
In this section, we propose our method to generate $l$-expressions of $l_{m}^{(i)},i=1,\cdots,m$ for an arbitrary kernel $G_{m}$. We begin with an example to illustrate the method. And we denote some functions in the description of the example. Then, a high-level description of the $l$-expressions generating algorithm is proposed according to these functions. Finally, we give details of these functions in general case and proofs of them.
\subsection{Illustration Example}
In Example \ref{l-ex_main_idea}, we use $l_{6}^{(4)}$ to illustrate the $l$-expressions generating algorithm.
Based on definition (\ref{def}), we get (\ref{eqn_dbl_x}). Actually, we omit known values $a_{1}=u_{1}+u_{2}+u_{3},a_{2}=u_{2},a_{3}=u_{3}$ in (5). We will add them in (\ref{final-l-expressions}). To omit known values at first and add them in final step, we call this function as \textsf{hide known values}.

Define $W(y_{3},y_{5}|u_{5})=W(y_{3}|u_{5})W(y_{5}|u_{5})$ in (\ref{extend1}) and (\ref{extend2}). Then it has
$l_{y_{3},y_{5}}\triangleq\frac{W(y_{3},y_{5}|0)}{W(y_{3},y_{5}|1)}=l_{3}l_{5}$. This function is called as \textsf{zero-variable-combine}.

By adding a mid term, we get (\ref{extend1}) and (\ref{extend2}) from (\ref{eqn_dbl_x}). $\boxtimes$ is the same as common multiple $\times$. Obviously, $(\ref{eqn_dbl_x})=(\ref{extend1})\boxtimes(\ref{extend2})$. We call (\ref{extend1}) and (\ref{extend2}) as \emph{sub-expressions}. $\boxtimes$ is specially used in separating sub-expressions.  It can be seen there are only one \emph{difference} in both (\ref{extend1}) and (\ref{extend2}). We call the function from (\ref{eqn_dbl_x}) to $(\ref{extend1})\boxtimes(\ref{extend2})$ as \textsf{extend method}.

With one difference property in (\ref{extend1}), we get the left part of (\ref{fundenmental1}). This is our key step and it is called \textsf{fundamental step}.

Firstly, let $W(y_{1}^{2}|u_{5}+u_{6})=W(y_{1}|\overline{u_{5}+u_{6}})W(y_{2}|u_{5}+u_{6})$ in (\ref{extend2}). It is \textsf{zero-variable-combine}. Then, we get the right part of (\ref{fundenmental1}) by defining $W(y_{1}^{2},y_{3},y_{5}|u_{6})=\sum_{u_{5}}W(y_{1}^{2}|u_{5}+u_{6})W(y_{3},y_{5}|u_{5})$. We call this function as \textsf{one-variable-combine}. With this function, we have
$l_{y_{1}^{2},y_{3},y_{5}}=l_{y_{1}^{2}}\diamond l_{y_{3},y_{5}}=l_{1}^{-1}l_{2}\diamond l_{3}l_{5}$.

The left part of (\ref{fundenmental2}) is obtained by doing $u_{5}=u_{6}$ and $u_{5}=u_{6}+1$ for each channel expressions in the numerator and denominator for the left part of (\ref{fundenmental1}), respectively. The right part of (\ref{fundenmental2}) is obtained by doing \textsf{zero-variable-combine} with defining $W(y_{1}^{2},y_{3},y_{5},y_{6}|u_{6})=W(y_{1}^{2},y_{3},y_{5}|u_{6})W(y_{6}|u_{6})$ and \textsf{fundamental step} in the right part of (\ref{fundenmental1}). Implementing \textsf{fundamental step} in the left part of (\ref{fundenmental2}) and doing $u_{6}=0$ and $u_{6}=1$ in both left and right of (\ref{fundenmental2}), we get (\ref{l-expressions}). Doing $l_{1}=l_{1}^{1-a_{1}},l_{2}=l_{2}^{1-a_{2}},l_{3}=l_{3}^{1-a_{3}}$ in (\ref{l-expressions}), we get (\ref{final-l-expressions}).

\subsection{High-Level Description of The Algorithm}
We denote \textsf{hide known values}, \textsf{zero-variable-combine}, \textsf{extend method} and \textsf{one-variable-combine} and \textsf{fundamental step} functions in the description of Example \ref{l-ex_main_idea}.

For a complete description of the $l$-expressions generating algorithm, we need two more functions \textsf{standard expression transform} and \textsf{symmetric expression transform}. The two functions are not necessary, but it can simplify the final $l$-expressions. Also, we use \textsf{two simplifications} to denote \textsf{zero-variable-combine} and \textsf{one-variable-combine}.

Based on these functions, we give a high-level description of $l$-expressions generating procedure in Algorithm \ref{algA}.
\begin{algorithm}
\caption{: Bit-channel likelihood expression generating procedure}
\label{algA}
\begin{algorithmic}
\STATE \textbf{input:} A kernel lower-triangular matrix $G_{m}$, index $i$ and channel output likelihood ratios $l_{1},\cdots,l_{m}$
\STATE \textbf{output:} Bit-channel likelihood expression $l_{m}^{(i)}$ as a function of $l_{1},\cdots,l_{m}$

\STATE//Algorithm starts from $l_{m}^{(i)}=\frac{\sum_{u_{i+1}^{m}}W(y_{1}|\textbf{u}_{1})\cdots W(y_{m}|\textbf{u}_{m})}
                {\sum_{u_{i+1}^{m}}W(y_{1}|\overline{\textbf{u}}_{1})\cdots W(y_{m}|\textbf{u}_{m})}$
\STATE\textbf{early processing:} implement \textsf{hide known values} and \textsf{standard expression transform} on $l_{m}^{(i)}$

 \WHILE{$i+1\leq m$}
 \FOR{each sub-expressions}
 \STATE implement \textsf{two simplifications} as much as possible
 \STATE implement \textsf{symmetric expression transform}
 \STATE implement \textsf{extending method}
 \STATE implement \textsf{fundamental step}
 \ENDFOR
 \ENDWHILE
\end{algorithmic}
\end{algorithm}

In Algorithm \ref{algA}, every step is working on the result from its previous step. After a \emph{while} \emph{loop} is finished, variables of expressions reduce at least $1$ (like left part of (9) to (10), $u_{5}^{6}$ to $u_{6}$). Then, the algorithm will stop after at most $m-i+1$ \emph{while} \emph{loops}.

\subsection{Details of functions}
\subsubsection{Hide known values}
Given a kernel $G_{m}$, let $G_{A}$ and $G_{B}$ be the submatrices of $G_{m}$ consisting of first $i-1$ rows and last $m-i+1$ rows, respectively.

Remember our basic task is to simplify (\ref{def}). Let $a_{j}=(u_{1}^{i-1}G_{A})_{j},j=1,\cdots,m$, we have
\[
W(y_{j}|(u_{1}^{m}G_{m})_{j})=W(y_{j}|a_{j}+(u_{i}^{m}G_{B})_{j}).
\]
Since $a_{j}$ are known values, we can \emph{replace} (\ref{def}) by following expression
\begin{equation}\label{hide}
l_{m}^{(i)}
   =\frac{\sum_{u_{i+1}^{m}}W(y_{1}|(u_{i,u_{i}=0}^{m}G_{B})_{1})\cdots W(y_{m}|(u_{i,u_{i}=0}^{m}G_{B})_{m})}
   {\sum_{u_{i+1}^{m}}W(y_{1}|(u_{i,u_{i}=1}^{m}G_{B})_{1})\cdots W(y_{m}|(u_{i,u_{i}=1}^{m}G_{B})_{m})}
\end{equation}

\begin{proposition}[Hide known values]
\label{hide_pro}
Assume we get the expression of $l_{m,(\ref{hide})}^{(i)}=f_{i}(l_{1},\cdots,l_{m})$ using (\ref{hide}) by Algorithm \ref{algA}. Then the real expression of $l_{m,(\ref{def})}^{(i)}$ using (\ref{def}) is
$l_{m,(\ref{def})}^{(i)}=f_{i}(l_{1}^{(1-2a_{1})},\cdots,l_{m}^{(1-2a_{m})})$. $\square$
\end{proposition}

\begin{IEEEproof}
For an arbitrary $j\in {1,\cdots,l}$,
\begin{align*}
  l_{j,(\ref{def})} & =\frac{W(y_{j}|a_{j}+(u_{i,u_{i}=0}^{m}G_{B})_{j}=0)}{W(y_{j}|a_{j}+(u_{i,u_{i}=1}^{m}G_{B})_{j}=1)} \\
  & =(\frac{W(y_{j}|(u_{i,u_{i}=0}^{m}G_{B})_{j}=0)}{W(y_{j}|(u_{i,u_{i}=1}^{m}G_{B})_{j}=1)})^{1-2a_{j}}= (l_{j,(\ref{hide})})^{1-2a_{j}}
\end{align*}
where $l_{j,(\ref{def})}$ and $l_{j,(\ref{hide})}$ mean $l_{j}$ using (\ref{def}) and (\ref{hide}), respectively.
\end{IEEEproof}

Based on Proposition \ref{hide_pro}, we implement the algorithm on (\ref{hide}) instead of (\ref{def}). After we get the final $l$-expressions, we replace $l_{j,(\ref{hide})}$ by
$(l_{j,(\ref{def})})^{1-2a_{j}}$ for each $j=1,\cdots,m$.
\subsubsection{Standard Expression Transform}
\begin{definition}[Standard expression]
A standard likelihood expression $l_{m}^{(i)}$ has following form
\begin{equation}\label{simple_bitchannel}
l_{m}^{(i)}=\frac{\sum_{u_{i+1}^{m}}\cdots W(y_{i+1}|u_{i+1})\cdots W(y_{m}|u_{m})}
                {\sum_{u_{i+1}^{m}}\cdots  W(y_{i+1}|u_{i+1})\cdots W(y_{m}|u_{m})};
\end{equation}
that is $\textbf{u}_{j}=u_{j},j=i+1,\cdots,m$.
\end{definition}

\begin{lemma}[Standard expression transform]
\label{SET}
For a likelihood expression $l_{m}^{(i)}$ defined by a lower triangular matrix $G_{m}$, it can be transformed to a standard expression.
\end{lemma}
\begin{IEEEproof}
First, we give an example of \emph{standard expression transform}.
Given a lower triangular matrix $G_{5}$
\[
G_{5}=\begin{array}{c}
         \\
         \\
        u_{3} \\
        u_{4} \\
        u_{5}
      \end{array}
\left(
        \begin{array}{ccccc}
          1 & 0 & 0 & 0 & 0 \\
          1 & 1 & 0 & 0 & 0 \\
          1 & 0 & 1 & 0 & 0 \\
          1 & 0 & 0 & 1 & 0 \\
          1 & 1 & 1 & 0 & 1 \\
        \end{array}
      \right).
\]
By doing linear transform in rows $3,4,5$ of $G_{5}$, we get
\[
\acute{G}_{5}=\begin{array}{c}
         \\
         \\
        t_{3} \\
        t_{4} \\
        t_{5}
      \end{array}
\left(
        \begin{array}{ccccc}
          1 & 0 & 0 & 0 & 0 \\
          1 & 1 & 0 & 0 & 0 \\
          1 & 0 & 1 & 0 & 0 \\
          1 & 0 & 0 & 1 & 0 \\
          0 & 1 & 0 & 0 & 1 \\
        \end{array}
      \right).
\]
Let $l_{5}^{(2)}$ and $\acute{l}_{5}^{(2)}$ are likelihood ratio expressions defined by $G_{5}$ and $\acute{G}_{5}$, respectively. Notice $\acute{l}_{5}^{(2)}$ is in \emph{standard expression transform}. Therefore, we only need to show $l_{5}^{(2)}=\acute{l}_{5}^{(2)}$. Since there is one-to-one correspondence between $u_{3}^{5}\in \{0,1\}^{3}$ and $t_{3}^{5}\in \{0,1\}^{3}$, it's easy to see $l_{5}^{(2)}=\acute{l}_{5}^{(2)}$ by the definition.

The above method can be easily generalized to any $l_{m}^{(i)}$ of a low triangular matrix $G_{m}$. Also, it's provide a procedure to implement \emph{standard expression transform}.
\end{IEEEproof}

In Lemma \ref{SET}, we suppose the kernel $G_{m}$ is a lower triangular matrix since all of optimal kernels given in \cite{Hisien-ping} are lower triangular matrices. In fact, we don't need lower triangular assumption in Lemma \ref{SET}. Because, given any $G_{m}$ and $l_{m}^{(i)}$, we always can transform $G_{C}$ ($G_{C}$ is the submatrix of $G_{m}$ consisting of last $m-i$ rows) to a lower triangular form with row transformation and column rearrangement.

\subsubsection{Fundamental Step}

\begin{lemma}[Fundamental step]
\label{fun_l}
Given a likelihood ratio expression with only one difference between the numerator and denominator such as
\begin{equation*}
l_{m}^{(i)}=\frac{\sum_{u_{i+1}^{m}}W(y_{1}|\textbf{u}_{1})W(y_{1}|\textbf{u}_{2})\cdots W(y_{m}|\textbf{u}_{m})}
                {\sum_{u_{i+1}^{m}}W(y_{1}|\overline{\textbf{u}}_{1})W(y_{1}|\textbf{u}_{2})\cdots W(y_{m}|\textbf{u}_{m})}
\end{equation*}
and assume that $\textbf{u}_{1}$ and $\textbf{u}_{m}$ contain $u_{i+1}$, then we have
\begin{equation}\label{fundamental}
  l_{m}^{(i)}=l_{1}\diamond\frac{\sum_{u_{i+2}^{m}}W(y_{2}|\textbf{x}_{2})\cdots W(y_{m}|\textbf{x}_{m})}
                {\sum_{u_{i+2}^{m}}W(y_{2}|\textbf{x}_{2})\cdots W(y_{m}|\overline{\textbf{x}}_{m})}
\end{equation}
where ${\textbf{x}}_{k}={\textbf{u}}_{k}+{\textbf{u}}_{1}$ if ${\textbf{u}}_{k}$ contains $u_{i+1}$; otherwise ${\textbf{x}}_{k}={\textbf{u}}_{k}$, $k=2,\cdots,m$ in the numerator of (\ref{fundamental}). In the denominator of (\ref{fundamental}), if ${\textbf{u}}_{k}$ contains $u_{i+1}$, ${\textbf{x}}_{k}={\overline{\textbf{x}}}_{k}$; otherwise ${\textbf{x}}_{k}={\textbf{x}}_{k}$.
\end{lemma}

\begin{IEEEproof}
Let $M=W(y_{1}|\textbf{u}_{2})\cdots W(y_{m}|\textbf{u}_{m})$ and define
\begin{align*}
  W(y_{2}^{m}|0) & \triangleq W(y_{2}^{m}|\textbf{u}_{1}=0)\triangleq \sum_{u_{i+1}^{m}}M\cdot\textbf{1}_{\textbf{u}_{1}=0}, \\
  W(y_{2}^{m}|1) & \triangleq W(y_{2}^{m}|\textbf{u}_{1}=1)\triangleq \sum_{u_{i+1}^{m}}M\cdot\textbf{1}_{\textbf{u}_{1}=1}.
\end{align*}
Then
\begin{align*}
  l_{m}^{(i)} &= \frac{\sum_{u_{i+1}^{m}}W(y_{1}|\textbf{u}_{1})M\textbf{1}_{\textbf{u}_{1}=0}+\sum_{u_{i+1}^{m}}W(y_{1}|\textbf{u}_{1})M\textbf{1}_{\textbf{u}_{1}=1}}
       {\sum_{u_{i+1}^{m}}W(y_{1}|\overline{\textbf{u}}_{1})M\textbf{1}_{\textbf{u}_{1}=0}+\sum_{u_{i+1}^{m}}W(y_{1}|\overline{\textbf{u}}_{1})M\textbf{1}_{\textbf{u}_{1}=1}} \\
   &= \frac{W(y_{1}|0)W(y_{2}^{m}|0)+W(y_{1}|1)W(y_{2}^{m}|1)}
               {W(y_{1}|1)W(y_{2}^{m}|0)+W(y_{1}|0)W(y_{2}^{m}|1)} \\
   &= l_{1}\diamond\frac{W(y_{2}^{m}|\textbf{u}_{1}=0)}{W(y_{2}^{m}|\textbf{u}_{1}=1)}\\
   &= l_{1}\diamond\frac{\sum_{u_{i+1}^{m}}W(y_{2}|\textbf{u}_{2})\cdots W(y_{m}|\textbf{u}_{m})\textbf{1}_{\textbf{u}_{1}=0}}
                {\sum_{u_{i+1}^{m}}W(y_{2}|\textbf{u}_{2})\cdots W(y_{m}|\textbf{u}_{m})\textbf{1}_{\textbf{u}_{1}=1}} \\
   & =l_{1}\diamond\frac{\sum_{u_{i+2}^{m}}W(y_{2}|\textbf{x}_{2})\cdots W(y_{m}|\textbf{x}_{m})}
                {\sum_{u_{i+2}^{m}}W(y_{2}|\textbf{x}_{2})\cdots W(y_{m}|\overline{\textbf{x}}_{m})}.
\end{align*}
\end{IEEEproof}

In Lemma \ref{fun_l}, we assume $\textbf{u}_{1}$ contains $u_{i+1}$. Then we have ${\textbf{x}}_{k}={\textbf{u}}_{k}+{\textbf{u}}_{1}$ if ${\textbf{u}}_{k}$ contains $u_{i+1}$; otherwise ${\textbf{x}}_{k}={\textbf{u}}_{k}$, $k=2,\cdots,m$ in the numerator of (\ref{fundamental}). In fact, we don't need this assumption. Assume $\textbf{u}_{1}=\{u_{i_{1}}+u_{i_{2}}+\cdots+u_{i_{t}}\}$, $1\leq i_{1}<i_{2}< \cdots < i_{t} \leq m$. We can choose any $u_{i_{j}},j\in \{1,2,\cdots,t\}$. Then it has ${\textbf{x}}_{k}={\textbf{u}}_{k}+{\textbf{u}}_{1}$ if ${\textbf{u}}_{k}$ contains $u_{t_{j}}$; otherwise ${\textbf{x}}_{k}={\textbf{u}}_{k}$, $k=2,\cdots,m$ in the numerator of (\ref{fundamental}). However, we always choose $u_{i_{1}}$ in the algorithm and it's good choice by experiments.

\begin{example}[Fundamental step for $l_{3}^{(1)}$]
\begin{align*}
  l_{3}^{(1)} & =\frac{\sum_{u_{2}^{3}}W(y_{1}|u_{2}+u_{3})W(y_{2}|u_{2})W(y_{3}|u_{3})}{\sum_{u_{2}^{3}}W(y_{1}|\overline{u_{2}+u_{3}})W(y_{2}|u_{2})W(y_{3}|u_{3})} \\
  &=l_{1}\diamond \frac{\sum_{u_{3}}W(y_{2}|u_{2})W(y_{3}|u_{3})\textbf{1}_{u_{2}+u_{3}=0}}{\sum_{u_{3}}W(y_{2}|u_{2})W(y_{3}|u_{3})\textbf{1}_{u_{2}+u_{3}=1}} \\
  &=l_{1}\diamond \frac{\sum_{u_{3}}W(y_{2}|u_{3})W(y_{3}|u_{3})}{\sum_{u_{3}}W(y_{2}|\overline{u_{3}})W(y_{3}|u_{3})}.
\end{align*}
\end{example}

By Lemma \ref{fun_l}, we reduce one variable from $u_{i+1}^{m}$ to $u_{i+2}^{m}$. So if there are still only one difference between the denominator and numerator of reduced expression (consider the left part in (8)), we can continue to implement Lemma1. Then we get the final likelihood expression after implementing $m-i$ times of Lemma \ref{fun_l}.

If there are more than one difference between numerator and denominator of expressions, we can define some mid-terms to extend expressions and make extending expressions have only one difference.

\begin{proposition}[Extend method]
\label{ex}
Given a likelihood ratio expression with two differences between the numerator and denominator such as
\[
l_{m}^{(i)}=\frac{\sum_{u_{i+1}^{m}}W(y_{1}|\textbf{u}_{1})W(y_{2}|\textbf{u}_{2})\cdots W(y_{m}|\textbf{u}_{m})}
                {\sum_{u_{i+1}^{m}}W(y_{1}|\overline{\textbf{u}}_{1})W(y_{2}|\overline{\textbf{u}}_{2})\cdots W(y_{m}|\textbf{u}_{m})}
\]
we have
\begin{align}
\label{extend_def1}
l_{m}^{(i)} & =\frac{\sum_{u_{i+1}^{m}}W(y_{1}|\textbf{u}_{1})W(y_{2}|\textbf{u}_{2})\cdots W(y_{m}|\textbf{u}_{m})}
                {\sum_{u_{i+1}^{m}}W(y_{1}|\overline{\textbf{u}}_{1})W(y_{2}|\textbf{u}_{2})\cdots W(y_{m}|\textbf{u}_{m})} \\
\label{extend_def2}
  &\boxtimes \frac{\sum_{u_{i+1}^{m}}W(y_{1}|\overline{\textbf{u}}_{1})W(y_{2}|\textbf{u}_{2})\cdots W(y_{m}|\textbf{u}_{m})}
                {\sum_{u_{i+1}^{m}}W(y_{1}|\overline{\textbf{u}}_{1})W(y_{2}|\overline{\textbf{u}}_{2})\cdots W(y_{m}|\textbf{u}_{m})}.
\end{align}
\end{proposition}

In proposition \ref{ex}, we divide the given $l_{m}^{(i)}$ into two part by operator $\boxtimes$  and it has only one difference in (\ref{extend_def1}) and (\ref{extend_def2}), respectively. $\boxtimes$ is the same as common multiple $\times$. $\boxtimes$ is specially used in \emph{extend method} and its priority is lower than $\times$.

It's easy to see that the extending method can be generalized to any number of differences. However, the cost is to increase the length of expression.

\subsubsection{Symmetric expression transform}
The length of final expression of $l_{m}^{(i)}$ is depend on the number of differences of $l_{m}^{(i)}$. Using the symmetric property of bit-channel, the number of differences can be reduced for a given $l_{m}^{(i)}$.
\begin{proposition}[symmetric property of bit-channel]
Given a bit-channel expression
\begin{equation*}
  W_{G_{m}}^{(i)}(\cdot|u_{i}=1)=\sum_{u_{i+1}^{m}}W(y_{1}|\overline{\textbf{u}}_{1})\cdots W(y_{i}|\overline{\textbf{u}}_{i})\cdots W(y_{m}|\textbf{u}_{m}),
\end{equation*}
and assume that only $\textbf{u}_{1}$ and $\textbf{u}_{i}$ contain $u_{i+1}$, we have
\begin{equation*}
  W_{G_{m}}^{(i)}(\cdot|u_{i}=1)= \sum_{u_{i+1}^{m}}W(y_{1}|\textbf{u}_{1})\cdots W(y_{i}|\textbf{u}_{i})\cdots W(y_{m}|\textbf{u}_{m}).
\end{equation*}
\end{proposition}

Proof is immediate and omitted. In this proposition, we change $u_{i+1}$ to $\overline{u}_{i+1}$. Actually, we can change all possible subsets of $u_{i+1}^{m}$. Let $(u_{i_{1}},\cdots,u_{i_{j}})$ denote a subset of $u_{i+1}^{m}$. We change it to $(\overline{u}_{i_{1}},\cdots,\overline{u}_{i_{j}})$. For $\textbf{u}_{k},k=1,\cdots,m$, if it contains odd number variables in $(u_{i_{1}},\cdots,u_{i_{j}})$, then $\textbf{u}_{k}$ becomes $\overline{\textbf{u}}_{k}$; otherwise it doesn't change.

\begin{proposition}[Symmetric expression transform]
\label{set}
Given a likelihood ratio expression
\[
l_{m}^{(i)}=\frac{\sum_{u_{i+1}^{m}}W(y_{1}|\textbf{u}_{1})\cdots W(y_{i}|\textbf{u}_{i})\cdots W(y_{m}|\textbf{u}_{m})}
                {\sum_{u_{i+1}^{m}}W(y_{1}|\overline{\textbf{u}}_{1})\cdots W(y_{1}|\overline{\textbf{u}}_{i})\cdots W(y_{m}|\textbf{u}_{m})},
\]
we use \emph{symmetric property of bit-channel} to all subsets of $u_{i+1}^{m}$ for denominator of $l_{m}^{(i)}$ and obtain $2^{m-i}$ equivalent likelihood ratio expressions. Assume that $\acute{l}_{m}^{(i)}$ has the least number of \emph{differences} among these expressions. Then we replace $l_{m}^{(i)}$ by $\acute{l}_{m}^{(i)}$.
\end{proposition}

Proposition \ref{set} describes a procedure to assure that $l_{m}^{(i)}$ has least number of difference. To do that, it need $2^{m-i-1}$ times tests. It's accepted for small kernel size. Actually, we just need to test one and two elements subsets of $u_{i+1}^{m}$ to acquire the least number of difference equivalent expression of $l_{m}^{(i)}$, up to kernel size $m=16$ by our experiments.

\subsubsection{Two Simplifications}
In this section, we propose two obvious ways to simplify the expressions.
\begin{proposition}[Zero-variable-combination]
Given a likelihood ratio expression as following
\[
l_{m}^{(i)} = \frac{\sum_{u_{i+1}^{m}}W(y_{1}|\textbf{u}_{1})W(y_{2}|\textbf{u}_{1})\cdots W(y_{m}|\textbf{u}_{m})}
                {\sum_{u_{i+1}^{m}}W(y_{1}|\textbf{u}_{1})W(y_{2}|\textbf{u}_{1})\cdots W(y_{m}|\overline{\textbf{u}}_{m})},
\]
we have
\[
l_{m}^{(i)} =\frac{\sum_{u_{i+1}^{m}}W(y_{1}^{2}|\textbf{u}_{1})\cdots W(y_{m}|\textbf{u}_{m})}
                {\sum_{u_{i+1}^{m}}W(y_{1}^{2}|\textbf{u}_{1})\cdots W(y_{m}|\overline{\textbf{u}}_{m})},
\]
where $W(y_{1}^{2}|\textbf{u}_{1})=W(y_{1}|\textbf{u}_{1})W(y_{2}|\textbf{u}_{1})$.
\end{proposition}

Then in the final likelihood expression, it has
$l_{y_{1}^{2}}=l_{1}l_{2}.$
\begin{proposition}[One-variable-combination]
Given a likelihood ratio expression as following
\[
l_{m}^{(i)} = \frac{\sum_{u_{i+1}^{m}}W(y_{1}|\textbf{u}_{1})W(y_{2}|\textbf{u}_{2})\cdots W(y_{m}|\textbf{u}_{m})}
                {\sum_{u_{i+1}^{m}}W(y_{1}|\overline{\textbf{u}}_{1})W(y_{2}|\textbf{u}_{2})\cdots W(y_{m}|\textbf{u}_{m})}
\]
and assume $\textbf{u}_{2}=u_{i+1}$, $\textbf{u}_{2}\subset \textbf{u}_{1}$ and $\textbf{u}_{2}\cap \textbf{u}_{k}=\emptyset$, $k=3,\cdots,m$. Then we have
\[
l_{m}^{(i)} =\frac{\sum_{u_{i+2}^{m}}W(y_{1}^{2}|\textbf{u}_{1}+\textbf{u}_{2})\cdots W(y_{m}|\textbf{u}_{m})}
                {\sum_{u_{i+2}^{m}}W(y_{1}^{2}|\overline{\textbf{u}_{1}+\textbf{u}_{2}})\cdots W(y_{m}|\textbf{u}_{m})}
\]
where $W(y_{1}^{2}|\textbf{u}_{1}+\textbf{u}_{2})=\sum_{\textbf{u}_{2}}W(y_{1}|\textbf{u}_{1})W(y_{2}|\textbf{u}_{2})$. $\square$
\end{proposition}

Then in the final likelihood expression, it has $l_{y_{1}^{2}}=l_{1}\diamond l_{2}.$
\begin{example}[One-variable-combination for $l_{3}^{(1)}$]
\begin{align*}
  l_{3}^{(1)} & =\frac{\sum_{u_{2}^{3}}W(y_{1}|u_{2}+u_{3})W(y_{2}|u_{2})W(y_{3}|u_{3})}{\sum_{u_{2}^{3}}W(y_{1}|\overline{u_{2}+u_{3}})W(y_{2}|u_{2})W(y_{3}|u_{3})} \\
  &=\frac{\sum_{u_{3}}W(y_{1}^{2}|u_{3})W(y_{3}|u_{3})}{\sum_{u_{3}}W(y_{1}^{2}|\overline{u}_{3})W(y_{3}|u_{3})}
\end{align*}
where $W(y_{1}^{2}|u_{3})=\sum_{u_{2}}W(y_{1}|u_{2}+u_{3})W(y_{2}|u_{2})$.
\end{example}

\section{Reduce Complexity by $W$-expressions}
In this section, we propose our methods to generate $W$-expressions of $W_{G_{m}}^{(i)}(\cdot|u_{i}),i=1,\cdots,m$ for an arbitrary kernel $G_{m}$. Firstly, we give an analysis about the complexity of $l$-expressions based SC decoder and show that $l$-expressions method is not accepted for larger kernels such as $m=16$. secondly, we analyse one drawback of $l$-expressions method and propose a $W$-expressions method to overcome the drawback. Then an example of $W$-expressions is given for making the method more clear. Finally, the complexity analysis of $W$-expressions based SC decoder is given and it contains our main achievement.

\subsection{Complexity Analysis of $l$-expressions}
Let $C_{l}(m)$ denote the average length of $l$-expressions for a kernel $G_{m}$. Actually $C_{l}(m)$ is the average number of sub-expressions. For a kernel $G_{m}$, the complexity of calculating a sub-expression is $O(m)$ and it needs to compute $mC_{l}(m)$ sub-expressions for the \emph{inside kernel calculation}.
Then the calculation cost of \emph{inside kernel calculation} is $ O(m^{2}\cdot C_{l}(m))$. Because it needs to implement $N\log_{m}^{N}/m$ times of \emph{inside kernel calculation}.
So the complexity of $l$-expressions based SC decoder is $O(C_{l}(m)\cdot mN\log_{m}^{N})$ for a general kernel $G_{m}$.


Table \ref{l_ex_com} gives the results of $C_{l}(m)$ by implementing Algorithm \ref{algA} for kernels up to $m=16$. It can be seen that the $l$-expressions method is good for small kernels such as $m\leq 10$. However, $C_{l}(m)$ increases very fast with kernel size $m$. Actually, $G_{16}$ is the first kernel which obtains significant advantages in terms of error performance compared with $G_{2}$. But $C_{l}(16)$ is about $2487$ times than $C_{l}(2)$. It means that $G_{16}$ based SC decoder is about $2487*16/(2*\log_{2}^{16})=4974$ times than the $G_{2}$ based SC decoder with the same block length. It can not be accepted.

\begin{table}[!ht]
\renewcommand{\arraystretch}{1.3}
\caption{$C_{l}(m)$ for different kernels}
\label{l_ex_com}
\centering
\begin{tabular}{|c|c|c|c|c|c|c|c|}
\hline
$m$ & 2 & 3 & 4 & 5 & 6 & 7 & 8\\
\hline
Exponent & 0.5 & 0.42 & 0.5 & 0.43 & 0.451 & 0.457 & 0.5 \\
\hline
$C_{l}(m)$ & 1.0 & 1.0 & 1.0 & 1.0 & 1.2 & 2.0 & 3.6 \\
\hline
9 & 10 & 11 & 12 & 13 & 14 & 15 & 16\\
\hline
0.461 & 0.469 & 0.477 & 0.492 & 0.488 & 0.494 & 0.5008 & 0.518\\
\hline
3.8 & 6 & 17 & 49 & 95 & 278 & 793 & 2487 \\
\hline
\end{tabular}
\end{table}

\subsection{Reduce Complexity by W-expressions}
It's shown in previous subsection that the complexity of $l$-expressions based SC decoder is unaccepted for large kernels. One problem is that we can not implement \textsf{two simplifications} in some cases because of the relation between numerator and denominator of $l_{m}^{(i)}$. For this reason, a method, called $W$-expressions, is proposed to further reduce the complexity of SC decoder by considering the numerator and denominator of $l_{m}^{(i)}$ separately.

In $W$-expressions, we focus on the following equation
\begin{equation}\label{w_def}
W_{G_{m}}^{(i)}(\cdot|u_{i})=\sum_{u_{i+1}^{m}}W(y_{1}|\textbf{u}_{1})\cdots  W(y_{m}|\textbf{u}_{m}).
\end{equation}

\begin{definition}
Let $B(y_{i})=(B(y_{i}|0),B(y_{i}|1))$. Define
\begin{align*}
B(y_{i})^{-1} &= (B(y_{i}|1), B(y_{i}|0)) \\
S(B(y_{i})) &= B(y_{i}|0)+B(y_{i}|1)\\
B(y_{i})\cdot B(y_{j}) &= (B(y_{i}|0)B(y_{j}|0), B(y_{i}|1)B(y_{j}|1))\\
B(y_{i})\diamond B(y_{j}) &= (B(y_{i}|0)B(y_{j}|0)+B(y_{i}|1)B(y_{j}|1) \\
 &\quad , B(y_{i}|0)B(y_{j}|1)+B(y_{i}|1)B(y_{j}|0))
\end{align*}
\end{definition}

\begin{lemma}[Fundamental step for $W$-expressions]
\label{f_w}
Given a channel expressions $W_{G_{m}}^{(\cdot)}(\cdot|u_{i})$ as (\ref{w_def}) and assume $\textbf{u}_{1}$ and $\textbf{u}_{m}$ contains $u_{1}$
, we have
\begin{multline}\label{fun_W}
  W_{G_{m}}^{(i)}(\cdot|u_{i})=\sum_{u_{i+1}^{m}}W(y_{1}|\textbf{u}_{1})\cdots W(y_{i}|\textbf{u}_{i})\cdots W(y_{m}|\textbf{u}_{m})\\
  =S(B(y_{1})\cdot (\sum_{x_{i+2}^{m}}W(y_{1}|\textbf{x}_{2})\cdots  W(y_{m}|\textbf{x}_{m}) \\
, \sum_{x_{i+2}^{m}}W(y_{1}|\textbf{x}_{2})\cdots  W(y_{m}|\overline{\textbf{x}}_{m})))
\end{multline}
where ${\textbf{x}}_{k}={\textbf{u}}_{k}+{\textbf{u}}_{1}$ if ${\textbf{u}}_{k}$ contains $u_{i+1}$; otherwise ${\textbf{x}}_{k}={\textbf{u}}_{k}$, $k=2,\cdots,m$ in the upper part of (\ref{fun_W}). In the lower part of (\ref{fun_W}), if ${\textbf{u}}_{k}$ contains $u_{i+1}$, ${\textbf{x}}_{k}={\overline{\textbf{x}}}_{k}$; otherwise ${\textbf{x}}_{k}={\textbf{x}}_{k}$.
\end{lemma}

\begin{IEEEproof}
Let $M=W(y_{1}|\textbf{u}_{2})\cdots W(y_{m}|\textbf{u}_{m})$ and define
\begin{align*}
  W(y_{2}^{m}|0) & \triangleq W(y_{2}^{m}|\textbf{u}_{1}=0)\triangleq \sum_{u_{i+1}^{m}}M\cdot\textbf{1}_{\textbf{u}_{1}=0}, \\
  W(y_{2}^{m}|1) & \triangleq W(y_{2}^{m}|\textbf{u}_{1}=1)\triangleq \sum_{u_{i+1}^{m}}M\cdot\textbf{1}_{\textbf{u}_{1}=1}.
\end{align*}
Then
\begin{align*}
  W_{G_{m}}^{(i)}(\cdot|u_{i}) &= \sum_{u_{i+1}^{m}}W(y_{1}|\textbf{u}_{1})M\textbf{1}_{\textbf{u}_{1}=0}+\sum_{u_{i+1}^{m}}W(y_{1}|\textbf{u}_{1})M\textbf{1}_{\textbf{u}_{1}=1} \\
   &= W(y_{1}|0)W(y_{2}^{m}|0)+W(y_{1}|1)W(y_{2}^{m}|1)\\
   &= S(B(y_{1})\cdot (W(y_{2}^{m}|0), W(y_{2}^{m}|1)))\\
   &= S(B(y_{1})\cdot (\sum_{x_{i+2}^{m}}W(y_{1}|\textbf{x}_{2})\cdots  W(y_{m}|\textbf{x}_{m})\\
   &\qquad\qquad\qquad, \sum_{x_{i+2}^{m}}W(y_{1}|\textbf{x}_{2})\cdots  W(y_{m}|\overline{\textbf{x}}_{m}))).
\end{align*}
\end{IEEEproof}

\begin{figure*}[t]
\begin{example}
\label{W-ex_main_idea}
\normalsize
\setcounter{mytempeqncnt}{\value{equation}}
\begin{align}
\label{W-def-64}
  W_{G_{6}}^{(4)}(y_{1}^{6},u_{1}^{3}|1) & = \sum_{u_{5}^{6}}W(y_{1}|\overline{u_{5}+u_{6}})W(y_{2}|u_{5}+u_{6})W(y_{3}|u_{5})W(y_{4}|\overline{u_{6}})W(y_{5}|u_{5})W(y_{6}|u_{6}) \\
\label{W-conb-zero1}
   & =\sum_{u_{5}^{6}}W(\overline{y_{1}},y_{2}|u_{5}+u_{6})W(y_{3},y_{5}|u_{5})W(\overline{y_{4}},y_{6}|u_{6}) \\
\label{W-conb-one}
  & =\sum_{u_{6}}W(\overline{y_{1}},y_{2},y_{3},y_{5}|u_{6})W(\overline{y_{4}},y_{6}|u_{6})  \\
\label{W-conb-zero2}
   & =\sum_{u_{6}}W(\overline{y_{1}},y_{2},y_{3},y_{5},\overline{y_{4}},y_{6}|u_{6}) \\
\label{W-fundenmental}
   &=S((B(y_{1})^{-1}\cdot B(y_{2}))\diamond (B(y_{3})\cdot B(y_{5}))\cdot (B(y_{4})^{-1}\cdot B(y_{6}))) \\
   W_{G_{6}}^{(4)}(y_{1}^{6},u_{1}^{3}|0) &= \sum_{u_{5}^{6}}W(y_{1}|u_{5}+u_{6})W(y_{2}|u_{5}+u_{6})W(y_{3}|u_{5})W(y_{4}|u_{6})W(y_{5}|u_{5})W(y_{6}|u_{6}) \\
\label{W-final-64-0}
   &= S((B(y_{1})\cdot B(y_{2}))\diamond (B(y_{3})\cdot B(y_{5}))\cdot (B(y_{4})\cdot B(y_{6})))
\end{align}
\hrulefill
\vspace*{4pt}
\end{example}
\end{figure*}

From (\ref{w_def}) to (\ref{fun_W}), we decompose $B(y_{1})$ from (\ref{w_def}) and the two remaining parts are in the same form as (\ref{w_def}). Therefore, we can use lemma \ref{f_w}  repeatedly and obtain an expression of $W_{G_{m}}^{(\cdot)}(\cdot|u_{i})$ as a function of $B(y_{1}),\cdots,B(y_{m})$.

The two remaining parts in (\ref{fun_W}) are called \emph{sub-expressions} for $W$-expressions. So the \emph{length} of $W$-expressions increases 1 after implementing \textsf{fundamental step} from (\ref{w_def}) to (\ref{fun_W}).

For $W$-expressions, \textsf{hiding known values} and \textsf{standard expression transform} are the same as $l$-expressions. In the final $W$-expressions, we have $B(y_{1}^{2})=B(y_{1})\cdot B(y_{2})$
for \textsf{zero-variables-combination} and $B(y_{1}^{2})=B(y_{1})\diamond B(y_{2})$
for \textsf{one-variables-combination}. It doesn't need \textsf{symmetric expression transform} and \textsf{extending method} for $W$-expressions.

Based on above, we give the $W$-expressions generating procedure in Algorithm \ref{algB}.
\begin{algorithm}
\caption{: $W$-expressions generating procedure}
\label{algB}
\begin{algorithmic}
\STATE \textbf{input:} A kernel matrix $G_{m}$, index $i$ and channel output $B(y_{1}),\cdots,B(y_{m})$
\STATE \textbf{output:} $W_{G_{m}}^{(i)}(\cdot|u_{i})$ as a function of $B(y_{1}),\cdots,B(y_{m})$

\STATE//Starts from $W_{G_{m}}^{(i)}(\cdot|u_{i})=\sum_{u_{i+1}^{m}}W(y_{1}|\textbf{u}_{1})\cdots  W(y_{m}|\textbf{u}_{m})$
\STATE\textbf{early processing:} implement \textsf{hiding known values} and \textsf{standard expression transform} on $W_{G_{m}}^{(i)}(\cdot|u_{i})$

 \WHILE{$i+1\leq m$}
 \FOR{each sub expressions}
 \STATE implement \textsf{two simplifications} as much as possible
 \STATE implement \textsf{fundamental step for $W$-expressions}
 \ENDFOR
 \ENDWHILE
\end{algorithmic}
\end{algorithm}

\subsection{Illustration Example for W-expressions}

In Example \ref{W-ex_main_idea}, we use $W_{G_{6}}^{(4)}(\cdot|u_{4})$ to illustrate $W$-expressions method.
Based on definition (\ref{def_bitchannel}), we get (\ref{W-def-64}). Then define $W(\overline{y_{1}},y_{2}|u_{5}+u_{6})=W(y_{1}|\overline{u_{5}+u_{6}})W(y_{2}|u_{5}+u_{6})$, $W(y_{3},y_{5}|u_{5})=W(y_{3}|u_{5})W(y_{5}|u_{5})$ and
$W(\overline{y_{4}},y_{6}|u_{6})=W(\overline{y_{4}}|u_{6})W(y_{6}|u_{6})$, we get (\ref{W-conb-zero1}).
Then we have
$B(\overline{y_{1}},y_{2})=B(y_{1})^{-1}\cdot B(y_{2})$, $B(y_{3},y_{5})=B(y_{3})\cdot B(y_{5})$ and $B(\overline{y_{4}},y_{6})=B(y_{4})^{-1}\cdot B(y_{6})$. They are \textsf{zero-variable-combine}.

(\ref{W-conb-one}) is obtained by defining
$W(\overline{y_{1}},y_{2},y_{3},y_{5}|u_{6})=\sum_{u_{5}}W(\overline{y_{1}},y_{2}|u_{5}+u_{6})W(y_{3},y_{5}|u_{5}).$ This is \textsf{one-variable-combine}. Then $B(\overline{y_{1}},y_{2},y_{3},y_{5})=B(\overline{y_{1}},y_{2})\diamond B(y_{3},y_{5})$.

Using \textsf{zero-variable-combine} again, we get (\ref{W-conb-zero2}). Then it has $B(\overline{y_{1}},y_{2},y_{3},y_{5},\overline{y_{4}},y_{6})=B(\overline{y_{1}},y_{2},y_{3},y_{5})\cdot B(\overline{y_{4}},y_{6})$.

Implementing \textsf{fundamental step for W-expressions}, we get (\ref{W-fundenmental}).

Similar to $W_{G_{6}}^{(4)}(\cdot|1)$, we give the $W$-expressions of $W_{G_{6}}^{(4)}(\cdot|0)$ in (\ref{W-final-64-0}).

Compared Example \ref{W-ex_main_idea} with Example \ref{l-ex_main_idea}, it can be seen that the $W$-expressions reduces the \emph{length} of expressions from 2 to 1. By our experiments, $W$-expressions offers significant advantages in terms of \emph{length} of expressions for larger kernels.

\begin{table}[htp]
\renewcommand{\arraystretch}{1.3}
\caption{$C_{W}(m)$ for different kernels}
\label{W_ex_com}
\centering
\begin{tabular}{|c||c|c|c|c|c|c|c|c|c|c|c|c|c|c|c|}
\hline
$m$ & 2 & 3  & 5 & 6 & 7 & 8 & 9 \\
\hline
$C_{W}(m)$ & 1 & 1 & 1 & 1 & 1.1 & 1.4 & 1.6 \\
\hline
$m$  & 10 & 11 & 12 & 13 &14 & 15 & 16 \\
\hline
$C_{W}(m)$ & 2.1 & 2.5 & 3 & 4 & 4.9 & 6.1 & 6.7\\
\hline
\end{tabular}
\end{table}

\subsection{Complexity Analysis of $W$-expressions}
In Example \ref{G6}, it can be seen that $l_{3}l_{5}$ just needs to calculate one time for $l_{6}^{(4)}$. We call this as \emph{inside expression simplification}. For the complexity analysis of $l$-expressions, we don't consider the \emph{inside expression simplification} since it makes no significant complexity reduction. However, it has significant complexity reduction by considering \emph{inside expression simplification} for $W$-expressions. For example, the length of $W_{16}^{(5)}$ is $512$. But we just need to calculate $16$ sub-expressions since other sub-expressions are the repetition of these $16$ sub-expressions.

Let $C_{W}(m)$ denote the average length of generated $W$-expressions for a general kernel $G_{m}$. Then the complexity of $W$-expressions based SC decoder is $O(C_{W}(m)\cdot mN\log_{m}^{N})$ for the $G_{m}$ based polar code. Table \ref{W_ex_com} gives the results of $C_{W}(m)$ by using Algorithm \ref{algB} for optimal kernels \cite{Hisien-ping} up to $m=16$. It means that the complexity of $W$-expressions based SC decoder is $O(m^{2}N\log N)$ when $m\leq 16$.

\begin{figure}[!h]
\centering
\includegraphics[width=7.0cm,height=6.0cm]{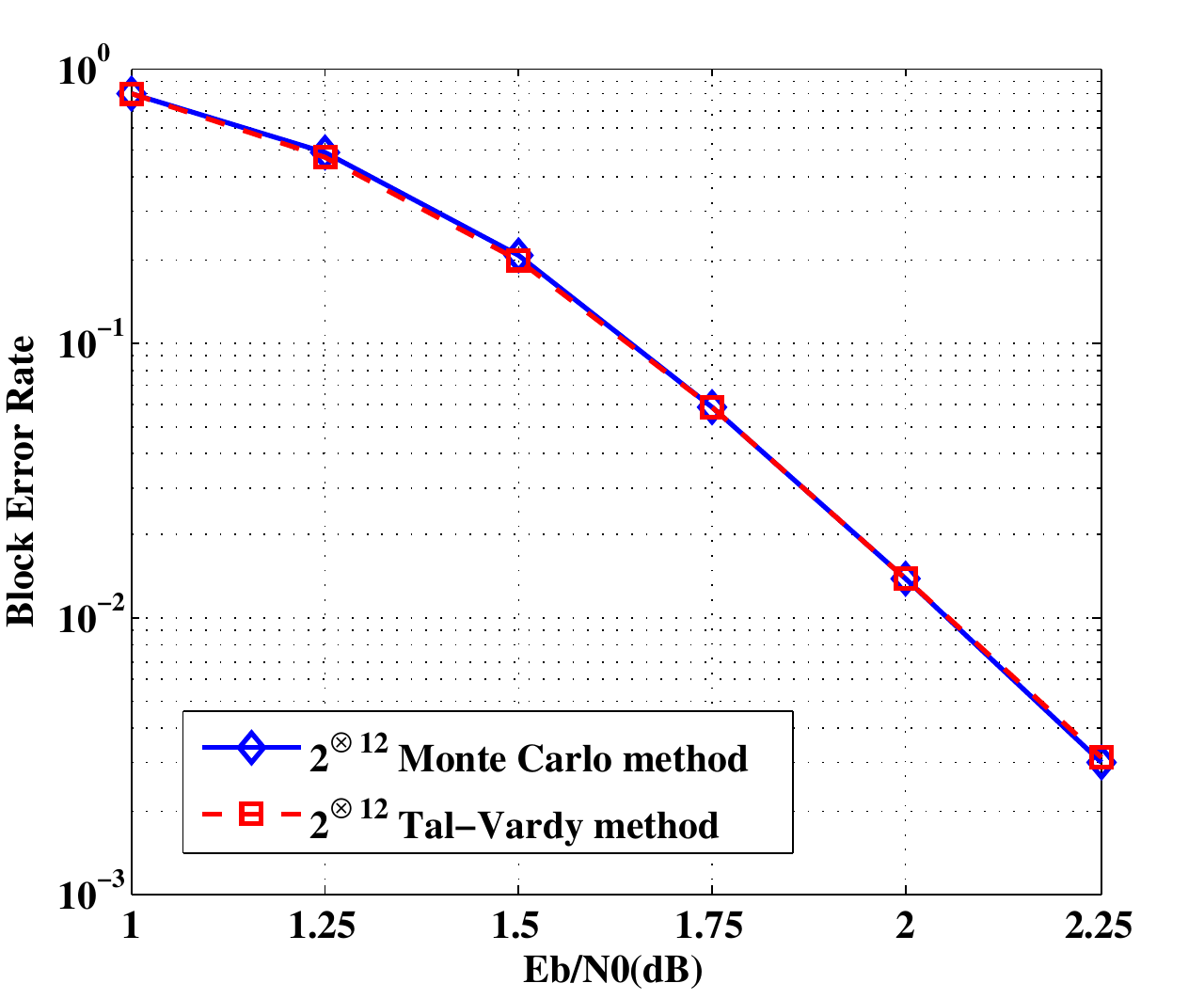}
\caption{Block-error-rate versus $E_{b}/N_{0}$ for SC decoding with $G_{2}^{\otimes 12}$ polar code on the BPSK-modulated Gaussian channel using Tal-Vardy method \cite{DE&UP} and Monte Carlo method \cite{Arikan} at Eb/N0$=2$dB. The code rate is 0.5.}
\label{fig-Tal-vs-MC}
\end{figure}
\section{Codes Construction}
Two methods are proposed to construct polar codes with high dimensional polar codes. One is Monte Carlo method \cite{Arikan}
, the other is Gaussian Approximation based density evolution (GA-DE) method  \cite{GA}.
\subsection{Monte Carlo method}
Ar{\i}kan [1, Section IX] provides a Monte Carlo approach for constructing polar codes. In Monte Carlo approach, it assume that all-zero codeword is transmitted. Firstly, for a bit-channel $W_{N}^{(i)}, i\in \{1,\cdots,m\}$, it assume that $u_{1}^{i-1}=0_{1}^{i-1}$ are known values. Then it uses SC decoder to evaluate the reliability of $W_{N}^{(i)}$. Finally, based on reliabilities of $W_{N}^{(i)}, i\in \{1,\cdots,m\}$, it chooses some best bit-channels as information set $\mathcal {A}$; that is the polar code.

Fig. \ref{fig-Tal-vs-MC} gives the comparison of error performances for $G_{2}^{\otimes 12}$ polar code which are constructed by the Monte Carlo method and Tal-Vardy method \cite{DE&UP}. Tal-Vardy method was considered the optimal construction method \cite{DE&UP}. It was shown that the Monte Carlo method achieves the same error performance as the Tal-Vardy  method. Therefore, it is conceivable that the Monte Carlo approach is an optimal method for constructing polar codes.

\begin{figure}[htb]
\centering
\includegraphics[width=7.0cm,height=6.0cm]{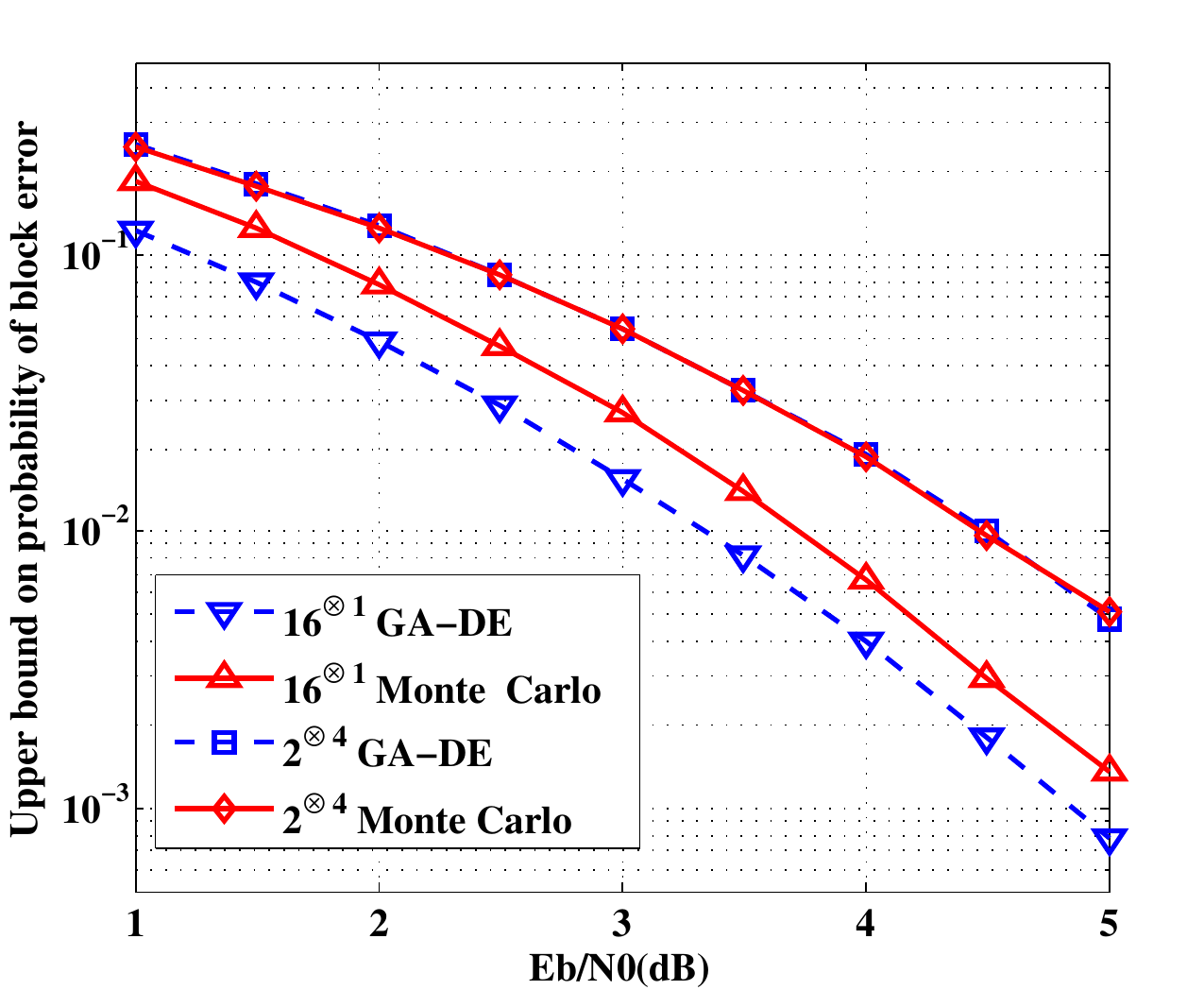}
\caption{$P_{fer}$ versus $E_{b}/N_{0}$ for SC decoding with $G_{2}^{\otimes 4}$ and $G_{16}^{\otimes 1}$ polar codes on the BPSK-modulated Gaussian channel using GA-DE method \cite{GA} and Monte Carlo method \cite{Arikan} at Eb/N0$=2$dB. The code rate is 0.5.}
\label{fig-GA-vs-MC}
\end{figure}

\subsection{Gaussian Approximation}
A first efficient construction of polar codes in general case which are based on density evolution \cite{DE} was made by Mori and Tanaka \cite{Mori-construction}. Then Trifinov demonstrated that polar codes can be efficiently constructed using GA-DE method \cite{Trifonov-GA}.

With $l$-expressions, it's straightforward to construct polar codes by using GA-DE method. However, polar codes constructed by GA-DE method become inaccurate as kernel size $m$ become larger by our experiments.

Let $W_{1},W_{2},\ldots,W_{N}$ be bit-channels and $P_{e}(W_{i})$ denote the probability of error on the $i$th bit-channel \cite{DE&UP}. Then a union bound on the frame error rate of polar codes is $P_{fer}\leq \sum_{i\in A}P_{e}(W_{i})$ where $\mathcal {A}$ is the \emph{information set} of the code \cite{Arikan}.

Fig. \ref{fig-GA-vs-MC} shows $P_{fer}$ vs. $E_{b}/N_{0}$ results under DE-GA and Monte Carlo methods. For the small block length $N=16$, it can be considered that Monte Carlo method is an accurate method for computing $P_{e}(W_{i})$. So Fig. \ref{fig-GA-vs-MC} confirms that polar codes constructed by GA-DE method become inaccurate as kernel size $m$ goes larger.

\ifCLASSOPTIONcaptionsoff
  \newpage
\fi

\end{document}